\documentclass[preprint,showpacs,preprintnumbers,amsmath,amssymb,showkeys]{revtex4}
\usepackage{graphicx}
\usepackage{dcolumn}
\usepackage{bm}

\begin{document}
\title{Investigation of the role of Ag in Pr$_{1-x}$Ag$_x$MnO$_3$ manganites}
\author{Hossein Ahmadvand}
\altaffiliation{Corresponding author} \email{ahmadvand@ph.iut.ac.ir}
\author{Hadi Salamati}
\author{Parviz Kameli}
\affiliation{Department of Physics, Isfahan University of Technology, Isfahan 84156-83111, Islamic Republic of Iran}
\date{\today}

\begin{abstract}
{X-Ray diffraction, ac susceptibility and electrical resistivity measurements were performed in order to investigate the effect of Ag substitution for Pr in the polycrystalline Pr$_{1-x}$Ag$_x$MnO$_3$ (0.0$\leq$x$\leq$0.25) manganites. The XRD results show that the samples crystallize in the O'-orthorhombic structure with a cooperative Jahn-Teller deformation. We give evidence that the Ag$^{+}$ is not substituted at the Pr$^{3+}$ site in the PrMnO$_{3}$ structure. The increase of $T_{C}$ up to 130 K is suggested to be due to further oxidation of manganite grains by the oxygen released from the metallic silver at high temperatures. No metal-insulator transition was observed in the samples.}
\end{abstract}


\maketitle
\section{Introduction}
Since the discovery of colossal magnetoresistive effects, the study of perovskite manganites Ln$_{1-x}$A$_{x}$MnO$_3$ (Ln=La$^{3+}$, Pr$^{3+}$,... ; A=Ca$^{2+}$, Sr$^{2+}$, Na$^{+}$, K$^{+}$, Ag$^{+}$,...) received considerable interest due to fascinating fundamental physics as well as due to their potential applications. Among these compounds, the La$_{1-x}$Ag$_{x}$MnO$_3$ is of great interest because of its high magnetoresistance at room temperature \cite{mr1, mr2}. The reported results indicate that the solubility of the Ag$^{+}$ in the LaMnO$_3$ is limited and the maximum solubility was reported to be at $x$=0.25 \cite{mr1}, or $x$=0.166 \cite{mr2}, above which  a magnetic perovskite phase and a nonmagnetic metallic silver will be present. Nevertheless, some authors believe that the monovalent Ag$^{+}$ is not substituted at the La$^{3+}$ site in the LaMnO$_3$ structure \cite{d,e}. The Nd$_{1-x}$Ag$_x$MnO$_3$ manganites have also been studied recently by Tang et al. \cite{ndag1} and Srivastava et al. \cite{ndag2}, who reported spin-glass-like behavior and enhanced low filed magnetoresistance in this system. Therefore, it is of interest to study the Ag doping in the other rare-earth manganites such as PrMnO$_3$.\\
The PrMnO$_{3}$ is an antiferromagnet with a Ne\'{e}l temperature, $T_{N}$, of 99 K \cite{prmno}. Studies of monovalent cations (Na, K) doped PrMnO$_{3}$ systems have been reported previously \cite{k1,k2,k3,h}. In contrast, to the best of our knowledge, there are no published results on the physical properties of the Pr$_{1-x}$Ag$_x$MnO$_3$ series. In the present study, we are interested in the systematic study of the Ag doping and understanding its role in the PrMnO$_3$ manganite. To investigate the Ag substitution at the Pr site at low doping contents, a Pr-deficient sample Pr$_{0.9}$MnO$_3$, was also prepared for comparison with the low Ag doped samples.

\section{Experimental}
Polycrystalline samples of Pr$_{1-x}$Ag$_x$MnO$_3$ ($x$=0, 0.05, 0.10, 0.15, 0.20, 0.25) and Pr$_{0.9}$MnO$_3$ manganites were prepared by the conventional solid state reaction route. Stoichiometric mixtures of high purity Pr$_6$O$_{11}$, Ag$_2$O and MnO$_2$ powders were calcined at 900 $^{\circ}$C for 20 h. The resulting powders were then pressed into pellets and annealed in air at 910 $^{\circ}$C for 20 h and then at 950 $^{\circ}$C for 24 h with intermediate grinding. The obtained powders were ground, pelletized, and sintered at 1110 $^{\circ}$C for 24 h, then slowly cooled to 945 $^{\circ}$C and finally cooled in the furnace to room temperature.\\
The structural characterization of the samples was carried out using a Philips X'pert X-ray diffractometer. Rietveld analysis of the XRD patterns were done with the "Profile matching with constant scale factor" calculation mode of the FULLPROF package \cite{a}. The real ($\chi'$) and imaginary ($\chi''$) parts of ac susceptibility were measured using a Lakeshore 7000 susceptometer. The dc resistivity measurements were performed by the standard four-probe method using a closed cycle helium refrigerator.

\section{Results and Discussion}
The XRD results show that all of the Pr$_{1-x}$Ag$_x$MnO$_3$ ($x$=0, 0.05, 0.10, 0.15, 0.20, 0.25) and Pr$_{0.9}$MnO$_3$ manganites crystallize in the \textit{Pnma} space group with O'-orthorhombic type of unit cell distortion ($b/\sqrt{2}<c<a$). The O'-orthorhombic distortions are characteristic for orbitally ordered manganites. Fig. 1 shows a typical XRD pattern along with Rietveld refinement for sample $x$=0.10. In addition to the major phase, the diffraction peaks of metallic silver (2$\theta$=38.2$^{\circ}$ and 44.4$^{\circ}$) and a weak peak corresponding to Mn$_3$O$_4$ (2$\theta$=36.2$^{\circ}$) were detected in the XRD patterns for $x\geq$0.15. As seen from the inset of Fig. 1, the intensity of silver peak increases with increasing of doping content for $x\geq$0.15. All samples fulfill the criterion $b/c$$<$$\sqrt{2}$  characteristic of a cooperative Jahn-Teller deformation \cite{c}. The doping content dependence of the distortion parameter (defined as $D=(a+c)/b\sqrt{2}$ \cite{b}) and unit cell volume is shown in Fig. 2. From this figure, it is seen that the unit cell volume and distortion parameter of samples $x$=0.05 and Pr$_{0.9}$MnO$_3$ are obviously larger than the corresponding values of other samples. The lattice parameters at room temperature, determined from Rietveld analysis, are summarized in Table I. As seen from the table, the difference between samples $x$=0.05 and Pr$_{0.9}$MnO$_3$ and the rest is mainly induced by lattice parameter $a$. In fact, comparing of the lattice parameters of sample $x$=0.05 with the other Pr$_{1-x}$Ag$_x$MnO$_3$ samples show that there is a elongation of MnO$_6$ octahedra along the $a$ direction accompanied with a counter effect of contraction along the $b$ direction. The lattice parameter $c$ does not change significantly relative to lattice parameters $a$ and $b$. Similar behaviors have been observed by Srivastava et al. \cite{ndag2} in the Nd$_{1-x}$Ag$_x$MnO$_3$ series.\\
The temperature dependence of the $\chi'$ for Pr$_{1-x}$Ag$_x$MnO$_3$ samples, measured in an ac field of 5 Oe with frequency of 111 Hz, is presented in Fig. 3. The $\chi'(T)$ curve for $x$=0.0 shows a peak at 83 K. Such behavior of the $\chi'(T)$ is connected with the onset of magnetic canted structure \cite{4}. In the undoped sample (i.e. PrMnO$_{3+\delta}$), the oxygen nonstoichiometry induces the presence of Mn$^{4+}$ ions with holes in the e$_g$ band. In the Pr$_{1-x}$A$_x$MnO$_3$ (A=Ca, Sr, Ba, Na, K) systems the canted arrangement is shown to exist when the content of the Mn$^{4+}$ ions is in the range of 5 - 15\% \cite{h}. For $x$=0.05, a ferromagnetic component appears in the $\chi'$ at around 119 K (Fig. 4), in addition to the low temperature peak at 81 K corresponding to canted magnetic structure.  A Curie-Weiss ($\chi=C/(T-\Theta$)) fit of the $\chi'(T)$ (not shown here) in the temperature range above T$_C$ and T$_{CANT}$ yields positive Curie-Weiss temperatures of $\Theta_1\sim$+118 K and $\Theta_2\sim$+82 K, indicating ferromagnetic interactions at the both transition temperatures of sample $x$=0.05. The susceptibility behavior of sample $x$=0.05 indicates a strong competition between the superexchange and double-exchange mechanisms. For the higher doping contents, $x\geq$0.10, the $\chi'(T)$ increases rapidly as $T_{C}$ is approached from above, passing through a maximum at a temperature somewhat below $T_{C}$. By using of the $d\chi'(T)/dT$, the $T_{C}$ values were found to be 126, 129, 130 and 127 K for $x$=0.10, 0.15, 0.20 and 0.25 respectively.
As can be seen in Fig. 4, the behavior of the $\chi'(T)$ for Pr$_{0.9}$MnO$_3$ is similar to that of sample $x$=0.05. Both of them display a ferromagnetic component at about 119 K followed by a peak at lower temperature corresponding to canted structure. Coexistence of ferromagnetic and canted phases in these two samples is consistent with the results in the Pr$_{1-x}$A$_x$MnO$_3$ (A=Ca, Sr, Ba, Na, K) systems. In theses systems, when the percentage of the Mn$^{4+}$ is in the range of 15 - 20\%, the canted arrangement coexist with a ferromagnetic arrangement \cite{h}.\\
The results of Refs. \cite{d} and \cite{e} show evidence that the Ag$^+$ is not substituted at the La$^{3+}$ site in the LaMnO$_{3}$ structure. Here for the Pr$_{1-x}$Ag$_x$MnO$_3$, the metallic Ag peaks appear in the XRD patterns for the higher doping contents (inset Fig. 1). To investigate the Ag substitution at the Pr site at low doping contents, it is useful to compare sample Pr$_{0.9}$MnO$_3$ with lowest doped sample i.e. Pr$_{0.95}$Ag$_{0.05}$MnO$_3$. Neutron diffraction study on the Pr$_{0.9}$MnO$_3$ showed the presence of a canted spin arrangement with a significant Jahn-Teller distortion \cite{f}. As can be seen in Fig. 2, the distortion parameter of sample Pr$_{0.9}$MnO$_3$ is obviously larger than the other samples. Significant difference between both the unit cell volume and distortion parameter of samples $x$=0.05 and Pr$_{0.9}$MnO$_3$ and other samples indicate the similarity between unit cell characteristics of these two samples. In other side, the $\chi'(T)$ data show similar magnetic behaviors for both of them (Fig. 4). Based on such similarities, one may expect that sample $x$=0.05 contains vacancies at the Pr site, similar to the Pr$_{0.9}$MnO$_3$. This means that the Ag$^+$ is not substituted at the Pr$^{3+}$ site in the crystal structure of the PrMnO$_3$. Therefore, similarity between samples $x$=0.05 and Pr$_{0.9}$MnO$_3$, appearance of metallic silver peaks in the XRD patterns for the higher doping contents and the published results on the La$_{1-x}$Ag$_x$MnO$_3$ manganites \cite{d,e} allow us to conclude that the Ag$^+$ can not substituted for the Pr$^{3+}$ in the PrMnO$_3$ structure. Therefore, it is located at the grain-boundaries in the form of metallic silver which is not detectable by conventional XRD at low doping contents. Similar to the La$_{1-x}$MnO$_3$ \cite{vac}, vacancy creation at the Pr site is limited and thus at a certain level of doping the Pr$_{1-x}$MnO$_3$ structure is no longer stable. Consequently, minor amount of a second phase due to Mn$_3$O$_4$ was detected in the XRD patterns for the higher doping contents.\\
So, why the ferromagnetic phase is enhanced with the increase of nominal Ag content?. Xu. et al. \cite{d} proposed that the decomposition of Ag$_{2}$O at high temperatures can efficiently increase the oxygen content of the nominal La$_{1-x}$Ag$_x$MnO$_3$ manganites. Oxygenation of manganite grains modify the ratio Mn$^{3+}$/Mn$^{4+}$ and lead to the enhancement of $T_{C}$. The Ag$_{2}$O decomposes at about 300 $^{\circ}$C \cite{d} to metallic Ag and O$_2$ in the first step of multi-step solid state reaction route. \emph{So, in order to better understand the nature of $T_{C}$ enhancement, one needs to consider the interaction between oxygen and metallic silver}.
The silver-oxygen system has been studied extensively in the past, due to its fundamental interests and technological applications. It has been known that oxygen is often present in metallic silver in multiple states including atomic, molecular and subsurface states. Such various oxygen species (were labeled as $O_{\alpha}$, $O_{\beta}$, $O_{\gamma}$,...) have different characteristics and some type of them may be stable and not desorbs up to above 900 K \cite{ag1,ag2,ag3}. Also, it was shown that one half of the lattice oxygen inside subsurface layers of Ag$_{2}$O is transformed to subsurface oxygen in metallic silver characterized by a quasimolecular structure, and thermal annealing up to 1000 K did not result in the removal of the residual subsurface oxygen \cite{ag4}. As a result, we proposed that the enhancement of $T_{C}$ up to 130 K is due to the oxygen release from the metallic silver at high temperatures or may be due to the oxygen release from the molten silver when the sintering temperature exceeds the melting point of silver (960 $^{\circ}$C). It is noteworthy that the Ag$_{2}$O acts as an in situ oxygen donor at high temperatures in the case of high-temperature superconductors \cite{donor,apl}. In addition, as reported by Kumar et al. \cite{ybco}, formation of silver oxide in the laser plume during pulsed laser evaporation of Ag and Ag-doped YBa$_2$Cu$_3$O$_{7-\delta}$ superconductor and La$_{0.7}$MnO$_{3-\delta}$ manganite targets leads to the incorporation of oxygen into the lattice by decomposition of the silver oxide at the surface of substrate. This process leads to the improvement of the physical properties of the films.\\
From the above explanations it is clear that the nominal Pr$_{1-x}$Ag$_x$MnO$_3$ manganites can be conveniently described by [Ag + self doped Pr$_{1-y}$MnO$_{3+\delta}$], where vacancy content $y$ will remain almost constant for the higher doping contents. The variation of unit cell volume and distortion parameter with doping content (Fig. 2) is consistent with the proposed model. According to Boujelben et al. \cite{vacpr}, the average radius of the vacancies at the Pr$^{3+}$ site is larger than the radius of Pr$^{3+}$. As discussed above, sample $x$=0.05 contains vacancies, thus its cell volume is larger than that of undoped sample. Sample Pr$_{0.9}$MnO$_3$ has larger cell volume than the sample $x$=0.05  because of its higher vacancy content. With further increasing of Ag content, the oxygen content of the samples increases and lead to the increase of Mn$^{4+}$ ions. As can be understood from Fig. 3, the content of Mn$^{4+}$ significantly increased in $x\geq0.10$. Thus, the decrease of cell volume with $x$ for $x\geq0.05$ can be explained by the increase of the content of Mn$^{4+}$ ($r_{Mn^{4+}}$=0.530 {\AA}, $r_{Mn^{3+}}$=0.645 {\AA}). For $x$=0.25, the cell volume slightly increased. The $T_C$ of this sample is also decreased. It seems that there is a correlation between cell volume and $T_C$ (Table I). The increase of lattice parameter $a$ and the decrease of lattice parameter $b$ for samples $x$=0.05 and Pr$_{0.9}$MnO$_3$, means the increase of distortion in these samples. In the other side, as observable in Fig. 2, the increase of Mn$^{4+}$ for the higher doping contents leads to the reduction of the distortion parameter. The distortion parameter is almost the same for $x\geq0.15$. It should be mentioned that the distortion parameter of an undistorted structure is 1. Our results are consistent with the results of Mantytskaya et al. \cite{6}, who reported that an antiferromagnetic-ferromagnetic phase transition occurs by increasing oxygen concentration in Pr$_{0.9}$MnO$_x$ system.\\
Temperature dependence of resistivity for all the Pr$_{1-x}$Ag$_x$MnO$_3$ manganites exhibit semiconducting behavior over the whole measured temperature range, which is similar to the Nd$_{1-x}$Ag$_x$MnO$_3$ system \cite{ndag1}. In Fig. 6, we show the temperature dependence of resistivity for sample $x$=0.20 in the temperature range of 95 - 270 K. As Fig. 6 shows, the resistivity curve is divided into two parts by a characteristic temperature, $T_{a}\approx$ 125 K. Such anomaly of resistivity in the vicinity of $T_{C}$ is associated with the onset of ferromagnetic phase. However, no metal-insulator transition was observed in the samples at around $T_{C}$, which indicate that the nominal Pr$_{1-x}$Ag$_x$MnO$_3$ manganites are ferromagnetic insulators.

\section{Conclusions}
The structural, magnetic and electrical properties of nominal Pr$_{1-x}$Ag$_x$MnO$_3$ compounds were investigated. We proposed that the enhancement of $T_C$ is not due to the substitution of the Ag$^+$ for the Pr$^{3+}$ and most likely originated from the oxygenation of the manganite structure by the oxygen released from the metallic silver at high temperatures.

\section*{Acknowledgement}
This work was supported by Isfahan University of Technology. The authors would like to thank Mr. Mohsen Hakimi for his help at the beginning of this work.

\newpage
\begin{figure}[t]
\includegraphics*[width=10cm]{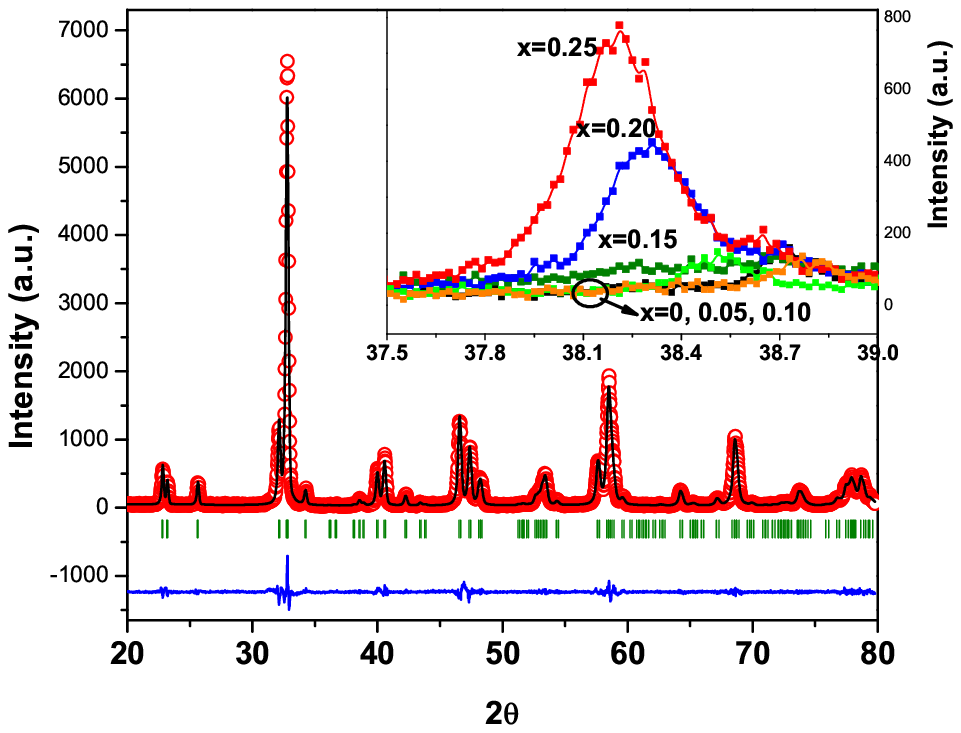}
\caption{The observed (circles) and calculated (solid line) XRD patterns of sample Pr$_{0.9}$Ag$_{0.1}$MnO$_3$. Inset shows the Ag peak of Pr$_{1-x}$Ag$_x$MnO$_3$ (0.0$\leq$x$\leq$0.25) samples.}
\end{figure}
\begin{figure}[t]
\includegraphics*[width=10cm]{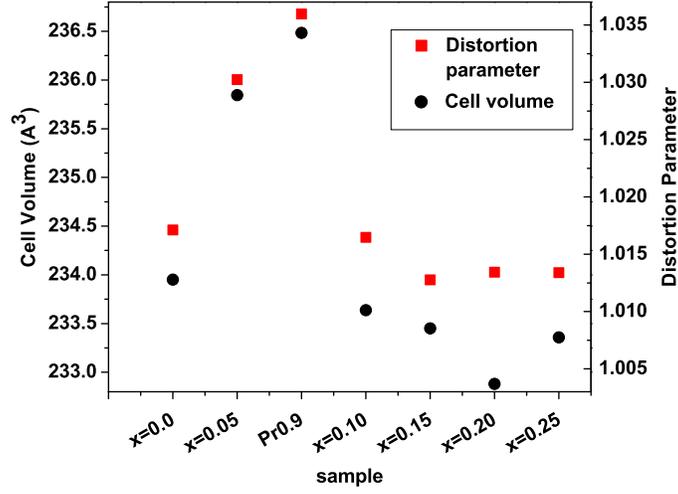}
\caption{Unit cell volume and distortion parameter of Pr$_{1-x}$Ag$_x$MnO$_3$ samples. The corresponding values of sample Pr$_{0.9}$MnO$_3$ are also added next to sample $x$=0.05 and indicated by Pr0.9.}
\end{figure}
\begin{figure}[t]
\includegraphics*[width=10cm]{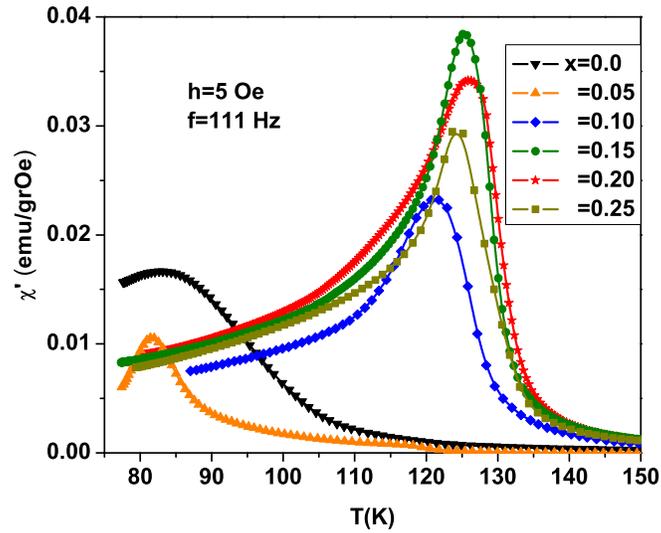}
\caption{The real part of ac susceptibility for Pr$_{1-x}$Ag$_x$MnO$_3$ (0.0$\leq$x$\leq$0.25) samples, measured in an ac field of 5 Oe with frequency of 111 Hz}
\end{figure}
\begin{figure}[t]
\includegraphics*[width=10cm]{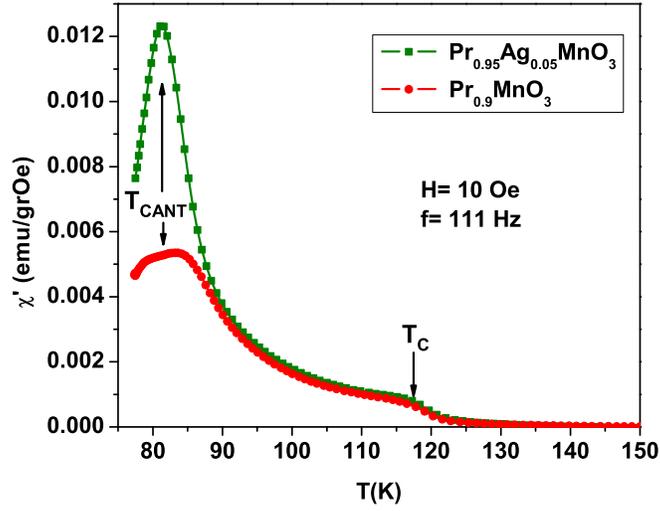}
\caption{The real part of ac susceptibility for samples Pr$_{0.95}$Ag$_{0.05}$MnO$_3$ and Pr$_{0.9}$MnO$_3$, measured in an ac field of 10 Oe with frequency of 111 Hz.}
\end{figure}
\begin{figure}[t]
\includegraphics*[width=10cm]{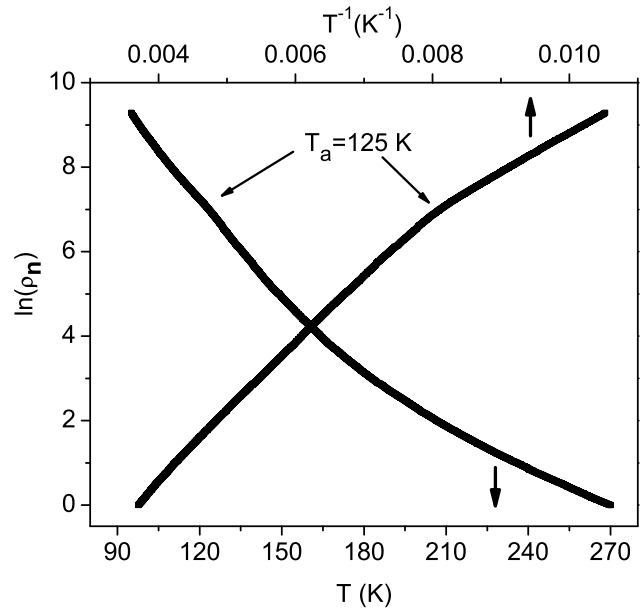}
\caption{Temperature dependence of the normalized resistivity ($\rho_n=\rho(T)/\rho(270)$) for sample $x$=0.20. For clarity, ln$\rho_n$ vs 1/T is also plotted.}
\end{figure}
\begin{table}
\caption{Summary of lattice parameters, cell volume and transition temperature for the polycrystalline samples of the Pr$_{1-x}$Ag$_x$MnO$_3$ and Pr$_{0.9}$MnO$_3$ manganites.}
\begin{tabular}{cddddd}
  \hline
  doping content (x) & a ({\AA}) & b ({\AA}) &  c ({\AA}) & V ({\AA}^3) & T_C(T_{CANT}) (K) \\\hline
  0.00  & 5.5816 & 7.6764 & 5.4602 & 233.95 & (83) \\
  0.05  & 5.6595 & 7.6320 & 5.4602 & 235.84 & 119 (81) \\
  0.10  & 5.5742 & 7.6761 & 5.4603 & 233.64 & 126 \\
  0.15  & 5.5570 & 7.6927 & 5.4610 & 233.45 & 129 \\
  0.20  & 5.5522 & 7.6831 & 5.4592 & 232.88 & 130 \\
  0.25  & 5.5585 & 7.6885 & 5.4604 & 233.36 & 127 \\
  Pr0.9 & 5.7017 & 7.6112 &	5.4493 & 236.48 & 119 (83) \\
  \hline
\end{tabular}
\end{table}

\section*{References}
\begin{thebibliography}{00}
\bibitem{mr1}
T. Tang, Q.Q. Cao, K.M. Gu, H.Y. Xu, S.Y. Zhang, Y.W. Du, Appl. Phys. Lett. {\bf 77} (2000) 723.
\bibitem{mr2}
L. Pi, M. Hervieu, A. Maignan, C. Martin, B. Raveau, Solid State Commun. {\bf 126} (2003) 229.
\bibitem{d}
Q.Y. Xu, R.P. Wang, Z. Zhang, Phys. Rev. B {\bf 71} (2005) 092401.
\bibitem{e}
V.L. Joseph Joly, P.A. Joy, S.K. Date, Appl. Phys. Lett. {\bf 78} (2001) 3747.
\bibitem{ndag1}
T.Tang, C.Tien, B.Y. Hou, Physica B {\bf 403} (2008) 2111.
\bibitem{ndag2}
S.K. Srivastava, S. ravi, J. Phys.: Condens. Matter. {\bf 20} (2008) 505212.
\bibitem{prmno}
J. Hemberger, M. Brando, R. Wehn, V.Yu. Ivanov, A.A. Mukhin, A.M. Balbashov, A. Loidl, Phys. Rev. B {\bf 69} (2004) 064418.
\bibitem{k1}
Z. Jir\'{a}k, J. Hejtm\'{a}nek, K. Kn\'{i}\v{z}eka, M. Mary\v{s}ko, E. Pollert, M. Dlouh\'{a}, S. Vratislav, R. Ku\v{z}el, M. Hervieu, J. Mag. Mag. Mater. {\bf 250} (2002) 275.
\bibitem{k2}
C. Shivakumara, M.S. Hegde, T. Srinivasa, N.Y. Vasanthacharya, G.N. Subbanna, N.P. Lalla, J. Mater. Chem., {\bf 11} (2001) 2572.
\bibitem{k3}
S. Zouari, A. Cheikh-Rouhou, P. Strobel, M. Pernet, J. Pierre, J. Alloys Comp. {\bf 333} (2002) 21.
\bibitem{h}
Z. Jir\'{a}k, J. Hejtm\'{a}nek, E. Pollert, M. Mary\v{s}ko, M. Dlouh\'{a}, S. Vratislav, J. Appl. Phys. {\bf 81} (1997) 5790.
\bibitem{a}
J. Rodriguez-Carvajal, Physica B {\bf 192} (1993) 55.
\bibitem{c}
J. T\"{o}pfer, J.B. Goodenough, J. Solid State Chem. {\bf 130} (1997) 117.
\bibitem{b}
C. Martin, A. Maignan, M. Hervieu, S. H\'{e}bert, A. Kurbakov, G. Andr\'{e}, F. Bour\'{e}e-Vigneron, J.M. Broto, H. Rakoto, B. Raquet, Phy. Rev. B {\bf 77} (2008) 054402.
\bibitem{4}
V. Dyakonov, F. Bukhanko, V. Kamenev, E. Zubov, S. Baran, T. Jaworska-Go{\l}\c{a}b, A. Szytua, E. Wawrzy\'{n}ska, B. Penc, R. Duraj, N. St\"{u}sser, M. Arciszewska, W. Dobrowolski, K. Dyakonov, J. Pientosa, O. Manus, A. Nabialek, P. Aleshkevych, R. Puzniak, A. Wisniewski, R. Zuberek, H. Szymczak, Phys. Rev. B {\bf 74} (2006) 024418.
\bibitem{f}
A. Mu\~{n}oz, J.A. Alonso, M.J. Mart\'{i}nez-Lope, M.T. Fern\'{a}ndez-D\'{i}az, Solid State Commun. {\bf 113} (2000) 227.
\bibitem{vac}
P.A. Joy, C. Raj Sankar, S.K. Date, J. Phys.: Condens. Matter {\bf 15} (2003) 3985.
\bibitem{ag1}
J.-H. Wang, W.-L. Dai, J.-F. Deng, X.-M. Wei, Y.-M. Cao, R.-S. Zhai, Appl. Surf. Sci. {\bf 126} (1998) 148.
\bibitem{ag2}
X. Bao, M. Muhler, Th. Schedel-Niedrig, R. Schl\"{o}gl, Phys. Rev. B {\bf 54} (1996) 2249.
\bibitem{ag3}
G.I.N. Waterhouse, G.A. Bowmaker, J.B. Metson, Appl. Surf. Sci. {\bf 214} (2003) 36.
\bibitem{ag4}
A.I. Boronin, S.V. Koscheev, O.V. Kalinkina, G.M. Zhidomirov, React. Kinet. Catal. Lett. {\bf 63} (1998) 291.
\bibitem{donor}
W.-H. Lee, Y. Abe, E. Inukai, J. Am. Ceram. Soc. {\bf 76} (1993) 849.
\bibitem{apl}
S. Sen, In-Gann Chen, C.H. Chen, D.M. Stefanescu, Appl. Phys. Lett. {\bf 54} (1989) 766.
\bibitem{ybco}
D. Kumar, S. Oktyabrsky, R. Kalyanaraman, J. Narayan, P.R. Apte, R. Pinto, S.S. Manoharan, M.S. Hegde, S.B. Ogale, K.P. Adhi, Mater. Sci. Eng. B {\bf 45} (1997) 55.
\bibitem{vacpr}
W. Boujelben, A. Cheikh-Rouhou, J. Pierre, J. C. Joubert, J. Alloy Comp. {\bf 315} (2001) 68.
\bibitem{6}
O.S. Mantytskaya, I.M. Kolesova, I.O. Troyanchuk, H. Szymczak, V.A. Sirenko, V.V. Eremenko, Low Temp. Phys. {\bf 32} (2006) 665.
\end{thebibliography}
\end{document}